\documentstyle[preprint,eqsecnum,aps,psfig]{revtex}

\begin{document}

\tightenlines
\draft
\preprint{OU-HET 329, hep-th/9910092}

\title{On the thermodynamics of large $N$ noncommutative super
Yang-Mills theory}

\author{Rong-Gen Cai\footnote{Email address: cai@het.phys.sci.osaka-u.ac.jp}
and Nobuyoshi Ohta\footnote{Email address: ohta@phys.sci.osaka-u.ac.jp}}
\address{Department of Physics, Osaka University, Toyonaka,
Osaka 560-0043, Japan}

\maketitle

\begin{abstract}

We study the thermodynamics of the large $N$ noncommutative super Yang-Mills
theory in the strong 't Hooft coupling limit in the spirit of AdS/CFT
correspondence. It has already been noticed that some thermodynamic
quantities of near-extremal D3-branes with NS $B$ fields, which are dual
gravity configurations of the noncommutative ${\cal N}$=4 super Yang-Mills
theory, are the same as those without $B$ fields. In this paper, (1) we
examine the $\alpha'^3 R^4$ corrections to the free energy and find that
the part of the tree-level contribution remains
unchanged, but the one-loop and the non-perturbative D-instanton corrections
are suppressed, compared to the ordinary case.  (2) We consider the
thermodynamics of a bound state probe consisting of D3-branes and D-strings
in the near-extremal D3-brane background with $B$ field, and find the
thermodynamics of the probe is the same as that of a D3-brane probe in the
D3-brane background without $B$ field.  (3) The stress-energy tensor of
the noncommutative super Yang-Mills theory is calculated via the AdS/CFT
correspondence. It is found that the tensor is  not isotropic and its trace
does not vanish, which confirms that the super Yang-Mills is not conformal
even  in four dimensions due to the noncommutative nature of space. Our
results render further evidence for the argument that the large $N$
noncommutative and ordinary super Yang-Mills theories are equivalent not only
in the weak coupling limit, but also in the strong coupling limit.
\end{abstract}

\newpage
\section{Introduction}

The super Yang-Mills theory (SYM) on noncommutative spaces is
a natural generalization of the SYM on the ordinary commutative spaces.
Such a noncommutative SYM has been found to arise naturally in a
certain limit of string theory with NS $B$
fields~\cite{Connes,Douglas,Ardalan,Sheikh,CHU,Seiberg}. The spirit of the
AdS/CFT correspondence~\cite{Mald,Itzhaki,Gubser,Witten} leads one to
try to find out the supergravity dual of the noncommutative SYM.
Recently Hashimoto and Itzhaki~\cite{Hash}, and Maldacena and
Russo~\cite{Mald1} constructed independently the supergravity dual
configurations of the noncommutative SYM's, which are the decoupling limits
of D-brane solutions with NS $B$ fields. Some of the latter have been
also constructed in~\cite{Russo,Breck,Lu} before~\footnote{These and more
general solutions are also discussed in~\cite{OZ} in IIA, IIB and $d=11$
supergravities.}.
The supergravity dual of the noncommutative SYM can also be constructed by
using the relationship between the open string moduli and closed string
moduli~\cite{Li99}. In this construction, the only input is a simple form
of the running string tension as a function of energy.

In the AdS/CFT correspondence, of particular interest is the D3-brane
solution. Its decoupling limit has the structure $AdS_5 \times S^5$, and
the type IIB string theory on this background is supposed to be dual to the
four-dimensional ${\cal N}$=4 SYM in the large $N$ and strong 't Hooft
coupling limit. At finite temperature the theory is described by the
near-extremal D3-brane configuration~\cite{Mald,Witten1}. According to the
AdS/CFT correspondence, the decoupling limit of D3-brane solutions with $B$
fields is supposed to be the dual gravity description of the noncommutative
SYM in four dimensions~\cite{Hash,Mald1}. An interesting question then
arises: Are the total numbers of degrees of freedom the same for the
noncommutative and ordinary SYM's at any given scale?  On the weak 't Hooft
coupling side, according to the analysis of planar diagrams~\cite{BS},
the large $N$ noncommutative and ordinary SYM's are equivalent; the planar
diagrams depends on the non-commutativity parameter only through the
external momenta and the noncommutative effects can be seen in the non-planar
diagrams. Explicit perturbative calculations~\cite{AV} provide evidence to
this assertion. On the strong 't Hooft coupling side, Maldacena and
Russo~\cite{Mald1} have discussed the thermodynamics of near-extremal
D3-branes with $B$ fields and found that the entropy and other thermodynamic
quantities are the same as those of the corresponding D3-branes without $B$
fields~\footnote{This conclusion also holds for other D-branes with $B$
fields. For related discussions see refs.\cite{AO,BR,HO}.}. On this basis,
they argued that the total number of physical degrees of freedom of the
noncommutative SYM at any given scale coincides with the ordinary case.

In the present paper we would like to investigate further aspects of
thermodynamics of the noncommutative SYM from the supergravity side
and to compare them with the ordinary SYM cases. In Section II, we introduce
the black D3-brane solutions with NS $B$ fields and calculate some of
their thermodynamic quantities. Most of the results are known, but these
are needed for our discussions. This also serves to establish our
notation. In Section III we calculate the corrections from the higher
derivative terms ($\alpha'^3 R^4$) to the free energy of the noncommutative
SYM. To compare the results with the ordinary case, we use a T-duality
transformation to transform the D3-brane solution with $B$ field and a
varying dilaton to that with constant dilaton and $B$ fields.
In the latter configuration, we find that the contribution coming from the
tree-level term remains the same as that in the ordinary case, but the
contributions from the one-loop and non-perturbative D-instanton terms are
suppressed. This result is consistent with that in the weak coupling
limit \cite{BS}.

In Section IV we consider the
thermodynamics of a static bound state probe consisting of D3-branes and
D-strings in the background produced by near-extremal D3-branes
with $B$ field. According to the interpretation of the D-brane action,
the supergravity interaction potential between the probe and the source
D-branes can be interpreted as the contribution of massive states to the
free energy of SYM when the SYM is in the Higgs phase, and the distance
between the probe and the source can be regarded as a mass scale of the
SYM. We find that the thermodynamics of the bound state probe again
remains the same as that of a D3-brane probe in the near-extremal
D3-brane background without $B$ field. In Section V, we compute
the stress-energy tensor of the noncommutative SYM on the supergravity side.
As is already known, the thermal excitations of D3-branes without $B$ fields
are of the form of an ideal gas in four dimensions. The entropy of
near-extremal D3-branes can be accounted for by the ideal gas
model~\cite{Gubser1}; its stress-energy tensor is isotropic and its trace
vanishes~\cite{Myers}, which confirms that the SYM is conformally invariant
in four dimensions. Our result shows that the stress-energy tensor of the
noncommutative SYM is not isotropic and its trace does not vanish,
which reflects the fact that the noncommutative SYM is not conformal
even in four dimensions. Section VI is devoted to the summary of our results
and discussions.


\section{The black D3-brane solution with $B$ field and its thermodynamics}

The supergravity solution corresponding to D3-branes with a non-vanishing NS
$B$ field has been constructed in~\cite{Russo} and \cite{Lu}. The simplest
way to get the solution is to start with a D3-brane solution without $B$
field. First make T-duality along $x_3$ (the world-volume coordinates are
$x_0$, $x_1$, $x_2$ and $x_3$), which gives a D2-brane solution with a
smeared coordinate $x_3$, perform a rotation with an angle $\theta$ in
the $x_2$-$x_3$ plane and then T-dualize back on $x_3$. This procedure
yields the desired solution with a non-vanishing $B$ field along $x_2$ and
$x_3$ directions~\cite{Mald1}. The prescription is also applicable to the
black D3-brane solutions. The black D3-brane solution with $B$ field along
$x_2$ and $x_3$ directions can be written in the string metric as
\begin{equation}
\label{1s1}
ds^2 = H^{-1/2} [-f dx_0^2 + dx_1^2 + h(dx_2^2 + dx_3^2)]
 + H^{1/2} [f^{-1}dr^2 + r^2 d\Omega_5^2],
\end{equation}
where
\begin{eqnarray}
&& H= 1 + \frac{r_0^4 \sinh ^2 \alpha}{r^4}, \ \
f = 1 -\frac{r_0^4}{r^4},\ \ h^{-1} = H^{-1}\sin^2\theta +\cos^2\theta,
 \nonumber \\
&& B^{(1)}_{23}=\frac{\sin \theta}{\cos\theta}H^{-1}h,\ \ e^{2\phi}=g^2 h,
 \ \ B^{(2)}_{01}=(1-H^{-1})\sin \theta \coth\alpha /g,
 \nonumber \\
&& C_{0123}=(1-H^{-1})h \cos\theta \coth\alpha /g.
\end{eqnarray}
The D3-brane charge satisfies $R'^4 \cos\theta =4\pi g\alpha'^2 N_3$.
Here $R'^4= r_0^4 \sinh\alpha \cosh\alpha $, $N_3$ is the number of
coincident D3-branes, and $g=g_{\infty}$ is the asymptotic value of the
coupling constant. The solution interpolates between the black D-string
solution ($\theta=\pi/2$) with the smeared coordinates $x_2$ and $x_3$ and
the black D3-brane solution  without $B$ field ($\theta=0$). In fact
the solution describes a non-threshold bound state consisting of D3-branes
and D-strings due to the presence of the nonzero $B$ field \cite{Lu}.

Taking the decoupling limit~\cite{Mald1}
\begin{eqnarray}
\alpha' \to 0: && \ \ \tan \theta =\frac{\tilde{b}}{\alpha'},
 \ \ x_{0,1}=\tilde{x}_{0,1},\ \ x_{2,3}=\frac{\alpha'}{\tilde{b}}
 \tilde{x}_{2,3},  \nonumber \\
\label{1s3}
&& \ \ r=\alpha' R^2 u,\  \ \ r_0=\alpha'R^2 u_0,\ \ \ g =\alpha'\tilde{g},
\end{eqnarray}
where $\tilde{b}$, $u$, $u_0$, $\tilde{g}$, and $\tilde{x_{\mu}}$ kept fixed,
the solution (\ref{1s1}) becomes
\begin{equation}
\label{1s4}
ds^2 =\alpha' R^2 \left [u^2(-\tilde{f}d\tilde{x}_0^2 +d\tilde{x}_1^2)
 +u^2 \tilde{h} (d\tilde{x}_2^2 +d\tilde{x}_3^2)
 +\frac{du^2}{u^2 \tilde{f}} + d\Omega_5^2\right],
\end{equation}
where
\begin{equation}
\label{1s5}
\tilde{f}=1-u_0^4/u^4, \ \ \tilde{h}^{-1}=1+a^4u^4, \ \
a^2=\tilde{b}R^2,\ \  e^{2\phi}=\hat{g}^2 \tilde{h},\ \ \tilde{B}_{23}
 =\frac{\alpha'}{\tilde{b}}\frac{a^4u^4}{1+a^4 u^4},
\end{equation}
and $\hat{g}=\tilde{g}\tilde{b}$ is the value of the string coupling
in the IR and $R^4=4\pi \hat{g} N_3=2g^2_{\rm YM}N_3 \equiv \lambda$ is
the 't Hooft coupling constant of gauge theory.

Let us first discuss the extremal case $\tilde{f}=1$ in the solution
(\ref{1s4}). The solution (\ref{1s4}) reduces to the familiar product
spacetime $AdS_5\times S^5$ for $a=0$, while it deviates from the anti-de
Sitter space for $a\neq 0$. Thus, in the spirit of AdS/CFT correspondence
the solution (\ref{1s4}) is proposed to be the gravity dual of the
noncommutative SYM and the parameter $a$ reflects the noncommutative
nature of space. When $u \to 0$, the solution (\ref{1s4}) approaches
the $AdS_5\times S^5$, which corresponds to the IR regime of the
gauge theory. This is in agreement with the expectation that the
noncommutative SYM reduces to the ordinary SYM at long distances.

Next, for non-extremal solution (\ref{1s4}), just like the pure black
D3-brane case, the thermodynamics of the non-extremal solution (\ref{1s4})
should be equivalent to that of the noncommutative SYM in the large
$N$ and strong 't Hooft coupling limit. However, the solution (\ref{1s4})
is neither asymptotically flat nor asymptotically anti-de Sitter. Hence it
is difficult to calculate the energy excitation of the noncommutative
SYM directly from the solution (\ref{1s4}). To discuss the thermodynamics
of the noncommutative theory, we rather start with the black D3-brane
solution (\ref{1s1}). For our purpose, it is convenient to rewrite the
solution in the Einstein frame, which has the following form:
\begin{equation}
\label{1s6}
ds^2_{\rm E}= h^{-1/4}H^{-1/2}[-f dx_0^2 + dx_1^2 +h(dx_2^2 +dx_3^2)]
 +h^{-1/4}H^{1/2}[f^{-1} dr^2 +r^2 d\Omega_5^2].
\end{equation}
We can further make compactification of the D3-brane world-volume and then go
to the Einstein frame. From the resulting metric, we can easily obtain the
ADM mass $M$, Hawking temperature $T$ and entropy $S$ of the solution, which
are found to be
\begin{eqnarray}
&& M=\frac{5\pi^2r_0^4 V_3}{16g^2 G_{10}}\left (1+\frac{4}{5}\sinh^2
 \alpha \right), \nonumber \\
&& T=\frac{1}{\pi r_0 \cosh\alpha}, \nonumber\\
\label{1s7}
&& S=\frac{V_3\pi^3}{4 g^2 G_{10}}r_0^5 \cosh\alpha,
\end{eqnarray}
where $G_{10}=2^3 \pi^6  \alpha'^4$ is the gravitational constant
in ten dimensions and $V_3$ is the spatial volume of the world-volume
of the D3-brane. We are interested in comparing these thermodynamic
quantities with those of black D3-branes without $B$ fields. We have just
found that these quantities are independent of the parameter $\theta$.
Thus they are exactly the same as those without $B$
field~\cite{Cai}~\footnote{The results in~\cite{Cai} are for rotating
D3-branes. For a comparison, take $l=0$ in the corresponding quantities
in~\cite{Cai}.}. Furthermore, it is worth pointing out that it is
independent of the parameter $\theta$ so that these thermodynamic quantities
(\ref{1s7}) are also those of black D-string solution with two smeared
coordinates. For later use, let us note the relation between the
numbers of D3-branes and D-strings in the solution (\ref{1s1}). The charge
density of D3-branes is
\begin{equation}
Q_3 =\frac{\pi^2 r_0^4}{4g G_{10}}\cos\theta \sinh\alpha \cosh\alpha,
\end{equation}
 while the charge density of D-strings is
\begin{equation}
Q_1= \frac{\pi^2 r_0^4V_2}{4g G_{10}}\sin\theta \sinh\alpha \cosh\alpha,
\end{equation}
where $V_2$ is the area the rectangular torus spanned by the two smeared
coordinates $x_2$ and
$x_3$. Using the charge quantization rule we obtain the following relation
between the number $N_3$ of D-branes and the number $N_1$ of D-strings:
\begin{equation}
\label{relation}
\frac{N_1}{N_3}=\frac{V_2 }{(2\pi)^2 \alpha'}\tan\theta .
\end{equation}

In the decoupling limit, the relation (\ref{relation}) becomes
\begin{equation}
\label{N}
\frac{N_1}{N_3}=\frac{\tilde{V}_2}{(2\pi)^2 \tilde{b}},
\end{equation}
where $\tilde{V}_2 =\tilde{b}^2 V_2/\alpha'^2$ is the area of the torus
after rescaling. The excitation above the extremality of the
black D3-brane corresponds to a thermal state of the corresponding
SYM. Considering the limit (\ref{1s3}), from (\ref{1s7})
we have the energy $E$, temperature $T$ and entropy of the large $N$
noncommutative SYM in the strong coupling limit:
\begin{eqnarray}
&& E=\frac{3\pi^3 \tilde{V}_3 R^8 u_0^4}{ (2\pi)^7 \hat{g}^2}, \nonumber \\
&& T=\frac{u_0}{\pi}, \nonumber \\
\label{1s8}
&& S= \frac{4\pi^4 \tilde{V}_3 R^8 u_0^3}{(2\pi)^7\hat{g}^2},
\end{eqnarray}
where $\tilde{V}_3=\tilde{b}^2V_3/\alpha'^2$. Obviously these thermodynamic
quantities satisfy the first law of thermodynamics $dE=TdS$. The free
energy $F$ of the gauge theory, defined as $F=E-TS$, can be expressed
in terms of the temperature:
\begin{equation}
\label{1s9}
F=-\frac{\pi^2}{8} \tilde{V}_3 N_3^2 T^4.
\end{equation}
Of course, the free energy is also the same as that of ordinary
SYM~\cite{Gubser1,Gubser2}. This result is quite interesting, which
leads Maldacena and Russo~\cite{Mald1} to argue that at any given scale
the total number of degrees of freedom of the noncommutative
SYM coincides with the ordinary case in the large $N$ limit. No doubt it
would be of much interest to further investigate this result and try to see
if this is modified by any corrections. Motivated by this observation, we are
now going to compute the higher derivative term corrections to the free
energy of the noncommutative SYM.


\section{The $\alpha'^3 R^4 $ corrections to the free energy}

The black configuration (\ref{1s1}) is an exact solution of type IIB
supergravity, which is a low-energy approximation of superstring, keeping
only the leading contribution of massless states in the $\alpha'$ expansion.
The non-leading contributions from massive string states appear as corrections
to this low-energy action in the form of higher derivative curvature terms.
In type IIB supergravity the lowest correction can be symbolically written
as $\alpha'^3 R_{\mu\nu\rho\sigma}^4$, where $R_{\mu\nu\rho\sigma}$
represents the Riemann tensor of spacetime. The tree-level contribution of
the four-graviton amplitude to the effective action is \cite{Green}
\begin{equation}
S^{\rm tree}_{R^4}=\frac{\zeta(3)}{3\cdot 2^6\cdot 16\pi G_{10}}
   \int d^{10}x\sqrt{-g}\alpha'^3 e^{-2\phi}R^4,
\end{equation}
in the string frame. Exploiting the field redefinition
ambiguity \cite{GW} and noting that for the D3-branes without $B$ field, the
extremal background $AdS_5\times S^5$ is a conformally flat spacetime,
the corrections can be written in the Einstein frame
as~\cite{Gubser2,Banks}
\begin{equation}
\label{2s1}
I^{\rm tree}_{R^4}=-\frac{\gamma }{16\pi G_{10}}
     \int d^{10}x\sqrt{g}e^{-3\phi/2}
 \left [C^{hmnk}C_{pmnq}C_h^{\ rsp}C^q_{\ rsk}
 +\frac{1}{2} C^{hkmn}C_{pqmn}C_h^{\ rsp}C^q_{\ rsk}\right],
\end{equation}
where $\gamma= \zeta (3)\alpha'^3 /8$ and $C_{pqmn}$ denotes Weyl
tensor. Such corrections to the free energy of the ordinary
SYM on the three-torus $T^3$, the three-sphere $S^3$, and even on a hyperbolic
space $H^3$ have been calculated in~\cite{Gubser2,Paw,Li1,Land,Cald,HO2}.
In particular, for the large three-torus $T^3$ case the free energy correction
is \cite{Gubser2}
\begin{equation}
\delta F^{\rm tree}_{R^4}= -\frac{\pi^2}{8}N^2_3 \tilde{V}_3T^4
     \frac{15}{8}\zeta(3) \lambda ^{-3/2}.
\end{equation}
Thus the free energy including the correction is
\begin{equation}
F_1= F +\delta F_{R^4}^{\rm tree}
          =-\frac{\pi^2}{8}N^2_3 \tilde{V}_3 T^4 \left [ 1 +
       \frac{15}{8}\zeta(3) \lambda ^{-3/2}\right],
\end{equation}
so that the leading correction is positive. If one writes the total
free energy as
\begin{equation}
F_{\rm total} = -f(\lambda )\frac{\pi^2}{6} N^2_3 \tilde{V}_3 T^4,
\label{fen}
\end{equation}
for the large 't Hooft coupling $\lambda$, one has
\begin{equation}
\label{correction}
f(\lambda ) =\frac{3}{4} +\frac{45}{32}\zeta (3)
 \lambda ^{-3/2} + \cdots.
\end{equation}
It is expected that the interpolation function $f$ smoothly approaches $1$ in
the weak coupling limit ($\lambda \to 0$) \cite{Gubser2}.

In fact in the type IIB supergravity the one-loop and non-perturbative
D-instanton contributions are also of the form $R^4$ and of the same order
($\alpha'^3$). If one writes
\begin{equation}
\rho =\rho_1 + i \rho_2 = c^{(0)} +i\ e^{-\phi},
\end{equation}
where $c^{(0)}$ is the RR pseudoscalar, the effective action of the $R^4$ part
can be expressed as \cite{Green,Banks}
\begin{equation}
S^{IIB}_{R^4} =\frac{1}{3\cdot 2^5\cdot 16\pi G_{10}}\int d^{10}x\sqrt{-g}
     \alpha'^3 e^{-\phi/2}f_4(\rho,\bar{\rho})R^4,
\end{equation}
where $f_4$ is given by the nonholomorphic Eisenstein series,
\begin{equation}
f_4(\rho,\bar{\rho})=\sum_{(m,n)\ne (0,0)}\frac{\rho_2^{3/2}}{|m+\rho n|^3}.
\end{equation}
For the small string coupling which is required for the validity of the
supergravity description, the function $f_4$ can be expanded as \cite{Banks}
\begin{eqnarray}
&& e^{-\phi/2} f_4 \approx 2\zeta(3) e^{-2\phi} +\frac{2\pi^2}{3}
     \nonumber \\
&&~~~~~+ (4\pi)^{3/2}e^{-\phi/2}\sum_{M>0}Z_M M^{1/2}\left(
         e^{-2\pi M(e^{-\phi}+ic^{(0)})}+e^{-2\pi M(e^{-\phi}-ic^{(0)})}
     \right) \left( 1+{\cal O}(e^{\phi}/M)\right),
\end{eqnarray}
where $M$ runs over integers.
Here the first term gives the tree-level contribution, the second term gives
the one-loop contribution, and the remaining denotes the contribution of the
non-perturbative D-instantons. The coefficient $Z_M$ is defined as
\begin{equation}
Z_M \equiv \sum_{m|M}\frac{1}{m^2},
\end{equation}
where $m|M$ denotes that the sum is taken over the divisors of $M$.
Considering the contributions from the one-loop term and from D-instantons,
$2\zeta (3) \lambda^{-3/2}$ in eq.~(\ref{correction}) is replaced
by~\cite{Gubser2}
\begin{equation}
\label{rep1}
2\zeta(3) \lambda^{-3/2} \to 2\zeta(3)\lambda^{-3/2} +
 \frac{1}{24N^2_3}\lambda^{1/2} +\frac{1}{N^{3/2}_3}h(e^{-4\pi^2
          /g^2_{\rm YM}})(1+{\cal O}(g^2_{\rm YM})),
\end{equation}
where $h$ represents infinite series of instanton corrections. In particular,
the one-loop contribution to the entropy correction ($\delta S
= -\delta(\delta F)/\delta T$) is
\begin{equation}
\delta S_{\rm one}=\frac{5\pi^2}{256}\lambda^{1/2}\tilde{V}_3 T^3.
\end{equation}

In the effective low energy action of type IIB supergravity, except for the
$R^4$ terms, in the same order ($\alpha'^3$) there  exist other terms,
for instance, eight-derivative four-dilaton term \cite{Brodie}, supersymmtric
terms accompanying $R^4$ terms, and so on (see \cite{Gubser2,Sethi} and
references therein). For the ordinary SYM, however, those terms will not
make contributions since the dilaton is a constant and the 5-form field
strength is the same as that in the extremal background. For the
noncommutative SYM, namely the black D3-branes with nonvanishing $B$ field,
from (\ref{1s5}) we see that the dilaton is no longer a constant and hence
its derivative terms and other possible terms involving the derivatives
of dilaton and curvature tensors are expected to make contribution to the
free energy correction. Also other terms unknown so far might have potential
contributions in this order. Unfortunately, till now there has not been a
complete expression of the effective low energy action to the order
($\alpha'^3$), to the best of our knowledge. This makes it difficult to
evaluate the free energy correction of noncommutative SYM via the
supergravity description and to compare with the ordinary SYM
case.\footnote{In the earlier version of this paper, we calculated the free
energy correction from the term (\ref{2s1}) and found that it is always
less than the ordinary case. We thank Troels Harmark and Niels Obers for
raising a question on the validity of that calculation to compare with the
ordinary case. We also thank the referee for valuable comments on this
point.}

To resolve this difficulty, we will adopt the following approach to attain
insight into the ($\alpha'^3$) correction to the free energy of the
noncommutative SYM, rather than the usual way to evaluate the free
energy correction by substituting the unperturbed solution (\ref{1s4}) into
those terms of the order ($\alpha'^3$) all of which are not known exactly.
As is well known, the T-duality is a perturbative symmetry of full string
theories valid loop by loop. This symmetry holds in the low energy
supergravities as well. The low energy effective action remains unchanged
under the T-duality transformation. Therefore if one can transform the
solution (\ref{1s4}) with the varying dilaton to a solution with a constant
dilaton, one may get the free energy correction of the former via the latter.
Indeed it has been found that such a T-duality transformation exists.

Following \cite{Has}, defining
\begin{equation}
\mu = \frac{\tilde{V}_2}{(2\pi)^2 \alpha'}\left (\tilde{B}_{23}
  + i\sqrt{G_{22}G_{33}}\right),
\end{equation}
the relevant T-duality transformation is given by the $SL(2,Z)$ transformation
\begin{equation}
\mu \to \hat{\mu} =\frac{a\mu +b}{c\mu +d},
\end{equation}
where $ad-bc=1$. Acting this transformation to the solution (\ref{1s4})
yields
\begin{eqnarray}
\label{dsolu}
&& ds^2 =\alpha' R^2 \left[u^2 (-\tilde{f}d\tilde{x}_0^2 +d\tilde{x}_1^2
        +d\hat{x}_2^2 +d\hat{x}_3^2) +\frac{du^2}{u^2\tilde{f}}
        + d\Omega^2_5 \right], \\
&& e^{2\phi}=\frac{(2\pi)^4\tilde{g}^2\tilde{b}^4}{\tilde{V}_2^2},
\ \ \ \hat{B}_{23}=\frac{\alpha'}{\tilde{b}},
\end{eqnarray}
for $c=-1$ and $d=\tilde{V}_2/(2\pi)^2\tilde{b}$ when the latter is an
integer. Note from (\ref{N}) that $d=N_1/N_3$ must be a rational number.
If this is not an integer, after some steps of Morita equivalence
transformation as in \cite{Has}, one can reach a solution like (\ref{dsolu}).
The solution (\ref{dsolu}) is asymptotically of the structure
$AdS_5\times S_5$, completely the same as that describing the ordinary
SYM at finite temperature. Actually the solution (\ref{dsolu}) describes
a twisted ordinary SYM due to the presence of a constant NS $B$ field.
The ordinary SYM lives on a dual torus with area $\hat{V}_2=(2\pi)^4
\tilde{b}^2/\tilde{V}_2$ and its Yang-Mills coupling constant is
\begin{equation}
\hat{g}^2_{\rm YM}=\frac{(2\pi)^3\tilde{g}\tilde{b^2}}{\tilde{V}_2}
         =g^2_{\rm YM}\frac{(2\pi)^2\tilde{b}}{\tilde{V}_2}.
\end{equation}
This ordinary SYM is equivalent to the noncommutative SYM described by the
solution (\ref{1s4}) in the sense of the Morita equivalence \cite{Has}.
Note further that the number of D3-branes in (\ref{dsolu}) is
$\tilde{V}_2N_3/(2\pi)^2\tilde{b}=N_1$ according to (\ref{N}), rather than
$N_3$ in (\ref{1s4}). It is quite interesting to note that the area,
the Yang-Mills coupling constant and the rank of the gauge group of the
ordinary SYM in (\ref{dsolu}) are different from those of the noncommutative
SYM in (\ref{1s4}), but that the 't Hooft coupling constants for
both theories are the same
\begin{equation}
\hat{\lambda}=2 \hat{g}_{\rm YM}^2 N_3\frac{\tilde{V}_2}{(2\pi)^2\tilde{b}}
      =\lambda.
\end{equation}
Considering the spatial volume of world-volume in (\ref{dsolu}) is
\begin{equation}
\hat{V}_3 = \frac{(2\pi)^4\tilde{b}^2}{\tilde{V}_2^2}\tilde{V}_3,
\end{equation}
we conclude that the thermodynamics of the solution (\ref{dsolu}) is
the same as the one of the solution (\ref{1s4}), that is, the thermodynamics
of the noncommutative SYM because the Hawking temperature is unchanged and
$N_1^2 \hat{V}_3 = N_3^2\tilde{V}_3$. Indeed, the Morita equivalence
transformation will not change the thermodynamics of gauge field theory
\cite{Has}. Thus we expect
that the $\alpha'^3 R^4$ correction in (\ref{dsolu}) gives us the free energy
correction of the noncommutative SYM in the order ($\alpha'^3$). As just
mentioned above, the advantage to consider the solution (\ref{dsolu}), rather
than (\ref{1s4}) is that one does not have to worry about the contributions
from the derivative terms of dilaton and possible other terms since the
dilaton and the NS $B$ field are constants here.

Now it is easy to get the free energy correction of noncommutative
SYM from the ($\alpha'^3$) terms in the effective low energy action according
to the above consideration. It is obtained from that in the ordinary case
discussed in \cite{Gubser2}, eqs.~(\ref{fen}), (\ref{correction}) and
(\ref{rep1}) with the replacements of $g^2_{\rm YM}$ by $\hat{g}_{\rm YM}^2$,
$\tilde{V}_3$ by $\hat{V}_3$ and $N_3$ by $N_1=\tilde{V}_2N_3/(2\pi)^2
\tilde{b}$, respectively. Considering
the invariance of the 't Hooft coupling constant, we get the correction
function ({\ref{correction}) of the noncommutative SYM by replacing
\begin{eqnarray}
&& 2\zeta(3)\lambda^{-3/2} \to 2\zeta(3)\lambda^{-3/2} +
   \frac{(2\pi)^4\tilde{b}^2}{\tilde{V}_2^2}\frac{1}{24N^2_3}\lambda^{1/2}
   \nonumber \\
\label{rep2}
&&~~~~~+\left (\frac{(2\pi)^2\tilde{b}}{\tilde{V}_2}\right)^{3/2}
   \frac{1}{N_3^{3/2}}h\left (e^{-\tilde{V}_2/g^2_{\rm YM}\tilde{b}}\right)
  \left [ 1 +{\cal O}\left (\frac{(2\pi)^2\tilde{b}}{\tilde{V}_2}g^2_{\rm YM}
 \right )\right].
\end{eqnarray}
Comparing this with the ordinary case (\ref{rep1}), we see that the first
term is unchanged, but the remaining two terms are suppressed because
$(2\pi)^2\tilde{b}/\tilde{V}_2 \ll 1$. Recall that the first term comes from
the tree-level contribution, the second term from the one-loop contribution
and the third is the non-perturbative D-instanton contribution. This indicates
that in the strong 't Hooft coupling the large $N$ noncommutative and ordinary
SYM's are equivalent because the first term corresponds to the planar
diagrams and the second term to the non-planar diagrams. This result
is also consistent with the argument, which is made in the weak 't Hooft
coupling limit, that planar diagrams depend on the noncommutativity parameter
only through the external momenta and non-planar diagrams are generally
more convergent than their commutative counterparts \cite{BS}. The previous
and the present sections provide evidence of the equivalence between the large
$N$ noncommutative and ordinary SYM's. It would be interesting to accumulate
further evidence for this equivalence. In the next section we will do it
by studying the thermodynamics of a probe brane in the background (\ref{1s4}).

\section{The thermodynamics of a probe brane}

We know that the solution (\ref{1s1}) describes $N_3$ D3-branes coinciding
with each other. The configuration represents the noncommutative
SYM with gauge group $U(N_3)$ in the Higgs branch, in which the vevs of scalar
fields are zero. Therefore the thermodynamics given in Section II is the one
for noncommutative SYM in the Higgs branch. We now want to discuss the
thermodynamics of the noncommutative SYM in the Coulomb branch,
in which the vevs of some scalar fields do not vanish. Corresponding to
the Coulomb branch should be a multicenter configuration of D3-brane
solutions. One of the simplest cases is that $N$ parallel coinciding
D3-branes are separated along a single transverse direction by a distance
from a single D3-brane. The gauge symmetry is then broken from $U(N+1)$
to $U(N)\times U(1)$ and the distance can be regarded as a mass scale in
the gauge field. However, no stable, multicenter, non-extremal configurations
of D-branes have been known.\footnote{It is possible to have
non-extremal configurations for continuously distributed D-branes.
For D3-branes, see~\cite{Kraus,Sfetsos}, for example.} As an approximation,
one may consider the probe method. That is, we put an unexcited probe brane
in the background of other non-extremal D-branes and regard this as an
approximate multicenter solution. Such a method has been used recently
to study the thermodynamics of SYM in the Higgs
phase~\cite{Tseyt2,Kiritsis2,Land1,Cai1}.

Considering that the noncommutative and ordinary SYM's have
the same thermodynamics at a given scale in the Higgs branch, it would be
interesting to compare them also in the Coulomb branch. To this aim, in this
section, we investigate the thermodynamics of a probe in the non-extremal
D3-brane background. According to the interpretation of D-brane action,
the supergravity interaction potential between the probe and the
near-extremal D3-branes (as the source) can be interpreted as the
contribution of massive states to the free energy of gauge fields
in the large $N$ and strong 't Hooft coupling limit~\cite{Tseyt2}.
When NS $B$ field is present, the dynamics of a probe D3-brane is governed by
the following action:
\begin{equation}
\label{4s1}
S = -T_3 \int d^4x e^{-\phi}\sqrt{-\det (\hat{G} - \hat{B}^{(1)})}
 - T_3 \int \hat {C} - T_3 \int \hat{B}^{(2)} \wedge \hat{B}^{(1)},
\end{equation}
where $T_3=1/(2\pi)^3\alpha'^2 $ is the tension of D3-brane. In fact this is
a bound state probe consisting of D3-branes and D-strings. An explicit
evidence for this is the tension of the probe is $T_3\sqrt{1+\tan^2\theta}$.
We will see more evidences below.

 Substituting
the solution~(\ref{1s1}) into the probe action~(\ref{4s1}), one has
\begin{equation}
\label{3e2}
S = -\frac{T_3V_3}{g\cos\theta}\int d\tau H^{-1}[\sqrt{f} - 1 + H_0 - H],
\end{equation}
where we have subtracted a constant potential at spatial infinity and
\begin{equation}
H_0=1+ \frac{R'^4}{r^4}.
\end{equation}
In the extremal background where $f=1$, one can see from (\ref{3e2}) that
the static interaction potential between the probe and the source vanishes.
Note that the source is a non-threshold bound state consisting of D3-branes
and D-strings, and the static potential will no longer vanish unless the probe
is also the same bound state. In the non-extremal background, of course, the
static potential exists always.
In the decoupling limit (\ref{1s3}), we arrive at
\begin{equation}
\label{4s6}
F_p= \frac{\tilde{V}_3 N_3  u^4}{2\pi ^2}\left[
  \sqrt{1-\frac{u_0^4}{u^4}}-1 + \frac{u_0^4}{2u^4} \right],
\end{equation}
which agrees with the result in~\cite{Tseyt2} and~\cite{Kiritsis2} for a
D3-brane probe in the near-extremal D3-brane background without $B$
field.\footnote{There is a small difference between the probe free
energies in~\cite{Tseyt2} and~\cite{Kiritsis2}, which arises as follows.
In the decoupling limit, although $H_0 \approx \frac{1}{\alpha'^2 R^4 u^4}$,
and $H \approx \frac{1}{\alpha'^2 R^4 u^4}$, the difference $H_0-H$ does not
vanish, but gives a finite value $\frac{u_0^4}{2u^4}$. This is just the
additional term appearing in~\cite{Kiritsis2}. The additional term is
important in the interpretation of the probe free energy.} When the probe is
on the horizon of the source, the free energy of the probe is
\begin{equation}
F_p|_{u=u_0}=-\frac{\tilde{V}_3 N_3 u_0^4}{4\pi^2}
    =-\frac{\pi^2 \tilde{V}_3N_3T^4}{4}.
\end{equation}
Comparing with the free energy of the source (\ref{1s9}), we find that
\begin{equation}
F_p|_{u=u_0}=\frac{dF}{dN_3}.
\end{equation}
The number of D3-branes in the probe is $1$, so we may rewrite the
above equation as
\begin{equation}
F_p|_{u=u_0}\approx F(N_3 +1)-F(N_3),
\end{equation}
for a large $N_3$. This implies that from the point of view of
thermodynamics, the non-extremal D-branes live on the horizon because the
probe branes on the horizon can be viewed as a part of source branes.

In the low-temperature or
long-distance limit, expanding the free energy (\ref{4s6}) and using
$u_0=\pi T$, we get~\footnote{Note that there is a difference by a factor of
$R^2$ in the rescaling of $r$ and $r_0$ from the definitions in \cite{Tseyt2}
and \cite{Kiritsis2}.}
\begin{equation}
\label{4s7}
F_p=-\frac{\pi^2 \tilde{V}_3N_3T^4}{4} \sum_{n=1}\frac{(2n-1)!!}{2^n(n+1)!}
 \left(\frac{\pi T}{u}\right)^{4n},
\end{equation}
 This is consistent
with the expectation that, in the weak coupling and low-temperature limit,
the contributions of one- and two-loops are exponentially
suppressed~\cite{Tseyt2,Kiritsis2}}. The leading term is a three-loop
contribution.

In the high-temperature or short-distance limit, we have to use the
isotropic coordinates defined in (\ref{3s3}) below~\footnote{Note that there
is a difference by a factor of $\sqrt{2}$ in the definition of the coordinate
$\rho$  between this paper and \cite{Tseyt2,Kiritsis2}. If we use the
definition in \cite{Tseyt2,Kiritsis2}, the metric will not be
asymptotically flat as $\rho \to \infty$.}.  Defining the mass
scale $M=(\sqrt{2}\rho -u_0)$, we obtain
\begin{equation}
\label{4s8}
F_p=-\frac{\pi^2 \tilde{V}_3N_3T^4}{4} \frac{1}{(1 + M/\pi T)^4}.
\end{equation}
Expanding (\ref{4s8}) for the small $M/\pi T$ yields
\begin{equation}
\label{4s10}
F_p=-\frac{1}{4}\pi^2 \tilde{V}_3 N_3T^4 \left[ 1
 - 4 \left( \frac{ M}{\pi T}\right) + 10 \left(\frac{M}{\pi T}\right)^2
 - 20 \left( \frac{M}{\pi T} \right)^3 + \cdots \right].
\end{equation}
Let us compare this with the free energy in the weak coupling limit.
The one-loop free energy of the ${\cal N}$=4 SYM in the weak coupling has
the following high-temperature expansion~\cite{Tseyt2}:
\begin{equation}
\label{4s11}
F_M(T\gg M) =-\frac{1}{3}\pi^2 N_3 \tilde{V}_3 T^4 \left [1-3
 \left (\frac{M}{\pi T}\right)^2 +
 4\left(\frac{M}{\pi T}\right)^3 +\cdots \right].
\end{equation}
It is very similar to that for the strong coupling limit (\ref{4s10})
except the term $(M/\pi T)$ is absent in the weak coupling. In particular,
in the massless approximation keeping only the leading terms in (\ref{4s10})
and (\ref{4s11}), one may see that there is also the well-known difference
by a factor of $3/4$, which occurs in comparing the supergravity calculation
and weak coupling calculation of the entropy for the ${\cal N}$=4 SYM in
the Higgs branch~\cite{Gubser1,Gubser2}.

The main result of this section is that the static interaction potential
between a D3-brane probe with NS $B$ field in the background of D3-branes
with $B$ field is the same as that of a D3-brane probe in the corresponding
background without $B$ field. From the point of view of field theory, the
static potential comes from planar diagrams \cite{Chep}, which is more clear
from the viewpoint of open strings extended between the probe branes and source
branes. This further renders evidence that the large $N$ noncommutative
and ordinary SYM's are also equivalent in the strong coupling limit.

\section{The Stress-energy tensor of the noncommutative SYM}

In Section II we have seen that the entropy of the noncommutative
SYM is the same as that of the SYM at a given temperature scale or
energy scale. However, from the Einstein frame metric of (\ref{1s4})
we see that, when
$u \to \infty $, the area of the torus $x_2$, $x_3$ contracts,
while the radius of the $S^5$ expands. The contraction of area of the
torus is just compensated by the expansion of the volume of the $S^5$.
This seems to imply that there is a redistribution of the degrees of
freedom~\cite{Mald1}. To compare the distribution of thermal states
between the noncommutative SYM and ordinary SYM, it is enough to
calculate the stress-energy tensor of the noncommutative SYM on the
supergravity side.

For this purpose, we adopt the method developed by Myers~\cite{Myers} by
generalizing the ADM mass density formula of $p$-branes~\cite{Lu1}. The
stress-energy tensor for the $p$-brane world-volume can be expressed as
\begin{equation}
\label{3s1}
T_{ab}=\frac{1}{16\pi g^2 G_{10}}\int_{r\to \infty} d\Omega_{8-p}
   r^{8-p}n^i [ \eta _{ab}(\partial_ih^c_{\ c} +\partial_i h^{j}_{\ j}
  -\partial_j h^{j}_{\ i}) -\partial_ih_{ab}],
\end{equation}
where $n^i$ is a radial unit in the transverse subspace, while
$h_{\mu\nu}=g_{\mu\nu}-\eta_{\mu\nu}$ is the deviation of the
(Einstein frame) metric from that for flat space. The labels
$a,\ b=0,1,\cdots,p$ run over the world-volume directions, while
$i,\ j =1,2,\cdots, 9-p$ denote the transverse directions. In addition,
it should be reminded that the calculations in (\ref{3s1}) must be done
using asymptotically Cartesian coordinates.

Rewriting the Einstein metric (\ref{1s6}) in isotropic coordinates,
one has
\begin{equation}
\label{3s2}
ds^2_{\rm E} = h^{-1/4}H^{-1/2}[-fdx_0^2 + dx_1^2 + h(dx_2^2 + dx_3^2)]
 + h^{-1/4}H^{1/2}r^2 \rho^{-2}[d\rho^2 + \rho^2 d\Omega_5^2],
\end{equation}
where
\begin{equation}
\label{3s3}
r^2=\rho^2 \left (1+\frac{r_0^4}{4\rho^4}\right), \ \
\rho^2 =\frac{1}{2} \left ( r^2 +\sqrt{r^4-r_0^4}\right).
\end{equation}
Substituting (\ref{3s2}) into (\ref{3s1}) and setting $p=3$ yields
\begin{equation}
\label{3s4}
T_{ab} = \frac{\pi^3}{16\pi g^2 G_{10}}
  {\rm diag} \left[5r_0^4 + 4 \tilde{R}^4,
  - r_0^4 - 4 \tilde{R}^4,
  - r_0^4 - 4 \tilde{R}^4\cos^2\theta,
  - r_0^4 - 4 \tilde{R}^4\cos^2\theta \right],
\end{equation}
where $\tilde{R}^4=\sqrt{R'^8 +r_0^8/4}-r_0^4/2$.
The stress-energy tensor (\ref{3s4}) includes the contribution from the
extremal background, which must be subtracted from it in order to acquire
the required quantity. The contribution of the extremal background
can be obtained directly from (\ref{3s4}) by setting $r_0=0$:
\begin{equation}
\label{3s5}
(T_{ab})_{\rm ext.} = \frac{\pi^3}{16\pi g^2 G_{10}}
 {\rm diag}\left[4 R'^4, - 4 R'^4,
 - 4 R'^4\cos^2\theta, - 4 R'^4 \cos^2 \theta \right].
\end{equation}
Subtracting (\ref{3s5}) from (\ref{3s4}) and taking the near-extremal
limit, $\tilde{R}^4 \approx R'^4-r_0^4/2$, we finally get the
stress-energy tensor for the noncommutative SYM in the large $N$ and
strong coupling limit
\begin{equation}
\label{3s6}
(\triangle T)_{ab} = \frac{\pi^3 r_0^4}{16\pi g^2 G_{10}}
 {\rm diag}[3,\ 1,\ 2\cos^2\theta - 1,\ 2\cos^2\theta - 1],
\end{equation}
and its trace
\begin{equation}
\label{3s7}
\triangle T = - \frac{4\pi^3 r_0^4}{16\pi g^2 G_{10}} \sin^2 \theta.
\end{equation}
For $\theta=0$, eq.~(\ref{3s6}) reduces to the result for the
ordinary SYM, which is of the form of an ideal gas in 3+1 dimensions.
In that case, its trace is zero. This is in accordance with the fact that
the ordinary SYM is conformally invariant in four dimensions. On the other
hand, for $\theta \neq 0$, eq.~(\ref{3s6}) gives the stress-energy tensor
for the noncommutative SYM. In this case, the tensor is not isotropic and
its trace does not vanish. It reflects the fact that the noncommutative
SYM is not conformal even in four dimensions due to the noncommutativity
of space. In addition, we confirm that
the $T_{00}$ component of the stress-energy tensor ({\ref{3s6}) in the
decoupling limit indeed gives the energy density of the noncommutative
SYM given in (\ref{1s8}).


\section{Conclusions}

To summarize, we have investigated some aspects of thermodynamics
for the noncommutative SYM in the large $N$ and strong 't Hooft
coupling limit on the supergravity side, and compared them with the
ordinary case. Although the entropy and other
thermodynamic quantities of black D3-branes with NS $B$ fields are the
same as those without $B$ fields, the stress-energy
tensor of thermal excitations is different. For the ordinary SYM
the stress-energy tensor is of the form of an ideal gas in four
dimensions. It is isotropic and its trace is zero. On the other hand, for
the noncommutative SYM, the tensor is not isotropic and its trace
does not vanish, which confirms that the noncommutative SYM is not
conformally invariant even in four dimensions due to the noncommutative
nature of space. Note that in the solution (\ref{1s1}) the NS $B$ field
has component only in $x_2$, $x_3$ directions. This means that the
coordinates $x_2$ and $x_3$ are noncommutative, while $x_0$ and $x_1$ are
the ordinary commutative coordinates. One may consider more general
D3-brane solutions with both $B_{01}$ and $B_{23}$ components. We do not
expect that the stress-energy tensor will be isotropic in that case either.
The result is indeed so, and we have confirmed this by a very similar
calculation to that in section V.

We have considered the higher derivative term corrections in the order
($\alpha'^3$) to the free energy of the noncommutative SYM in the
section III.  Because there has not been a complete expression of the
low energy effective action of Type IIB supergravity to the order
($\alpha'^3$), to make sense of the calculation and to compare the case
of the ordinary case which has been investigated in \cite{Gubser2},
we transformed the near-extremal  D3-brane solution with varying dilaton
to a solution (\ref{dsolu}) with a constant dilaton by a T-duality
transformation. Those two solutions are equivalence in the sense of the
Morita equivalence. Using the latter solution, we have found that the
tree-level contribution is the same as the ordinary case, but the one-loop
and the non-perturbative D-instanton contributions are suppressed, compared
to the ordinary case. Note that the tree-level part corresponds to the planar
diagrams, and the one-loop part to the non-planar diagrams in the field
theory. This provides evidence that the large $N$ noncommutative and
ordinary SYM's are also equivalent in the strong 't Hooft coupling limit.

We have also studied the thermodynamics of a bound state probe consisting of
D3-branes and D-strings in the background produced by the black D3-branes
with $B$ fields and compared it with that of a D3-brane probe  in the
background produced by the black D3-branes without $B$ fields. In accordance
with the interpretation of the D-brane action, the free energy of a static
probe can be regarded as the contribution of massive states to the free
energy of noncommutative SYM in the Higgs phase and the distance between
the probe and the source can be explained as a mass scale in the gauge
theory. From the thermodynamics of the probe we have found that the
free energies for the ordinary and noncommutative cases agree.
In fact the dynamic interaction potential between the probe and the
source also agrees with the ordinary case. Because the interaction potential
comes from the planar diagrams from the point
of view of field theory, the agreement further suggests that the large $N$
noncommutative and ordinary SYM's are equivalent not only in the weak
coupling limit \cite{BS}, but also in the strong coupling limit.


\section* {Acknowledgments}

We would like to thank T. Harmark and N. Obers for their very helpful
correspondences. This work was supported in part by the Japan Society for
the Promotion of Science and by grant-in-aid from the Ministry of Education,
Science, Sports and Culture No. 99020.

\end{document}